# Microwave magnetic field imaging based on Rabi resonance with an alkali-atom vapor cell

Xiaochi Liu,[1, a)] Songbai Kang,[2] Zhenfei Song,[1, b)] Zhiyuan Jiang,[1] Wanfeng Zhang,[1] and Jifeng Qu[1]

[1]*Center for Advanced Measurement Science, National Institute of Metrology, Beijing, 100029, China*

[2]*Key Laboratory of Atomic Frequency Standards, Wuhan Institute of Physics and Mathematics, Chinese Academy of Sciences, Wuhan, 430071, China*

In this work, we demonstrate a microwave magnetic field imaging technique based on Rabi resonance with a cesium atom vapor cell. Rabi resonance signals are generated when atoms interact with a phase-modulated microwave (MW) field and are detected by a photodiode and camera. A low noise, high quantum efficiency camera is used to capture a series of frames for different phase modulation frequencies. Rabi frequencies of each spatial point in the field can be measured by scanning the frames. Thus, the strength of the MW magnetic field distribution is obtained by combing the Rabi frequencies, measured with camera pixels in the range of the probe laser beam. The simple architecture of this imaging setup holds great potential for the construction of compact/miniature MW field sensors for material testing, field imaging of MW components, and biomedical imaging.

Several quantum sensing techniques that use microwave (MW) fields have been demonstrated, such as MW electrometry based on Rydberg atoms;[1-5] MW magnetometry based on Rabi oscillation, with atomic vapor cell/cold atoms/diamond nitrogen-vacancy centers;[6-10] and MW magnetic field detection and standards with Rabi resonance signals, generated using a phase modulated MW signal.[11-16] In these atom-based MW sensing techniques, MW field strength is associated with Rabi frequency, which directly link parameter of MW field via atomic constants. Thus, the quantum sensing techniques of MW field detection have an advantage of high accuracy and are International System of Units (SI) – traceable compared with conventional MW detection techniques.[17-20]

In our earlier work, we developed a Rabi resonance-based MW magnetic field detection system and demonstrated its broadband sensing capability.[15-16] However, Rabi resonance

---

Author to whom correspondence should be addressed. Electronic mail: a) liuxch@nim.ac.cn; b) songzf@nim.ac.cn

signals in the experiment were detected with a low noise photodiode (PD), and the signals detected were averaged from the range of the probe laser beam. For some applications, such as near-field detection, monolithic MW integrated circuits and cavity testing, detection of explosives, and biomedical imaging,[21-25] a more detailed distribution of MW magnetic field imaging is required. Various imaging methods based on Rabi oscillation have been proposed, each with different operating mediums.[6-7, 10, 26]

Here we present a simple MW magnetic field imaging technique based on Rabi resonance generated with an alkali atom-buffer gas vapor cell. Similar to our previous work,[15-16] the MW magnetic field is detected using the Rabi resonance signal lineshape. The theoretical model of Rabi resonance is described by Camparo et al.[27-29] When atoms are resonant with a phase-modulated MW field, the atomic population in the excited state oscillates with an amplitude of $P_e$, which is given as:

$$P_e(t) = P_\alpha \sin(\omega_m t + \phi_\alpha) + P_\beta \sin(2\omega_m t + \phi_\beta), \qquad (1)$$

where $P_\alpha$ and $P_\beta$ are the oscillation amplitudes for phase modulations $\omega_m$ and $2\omega_m$, respectively. The two different oscillation amplitudes are termed $\alpha$ and $\beta$ Rabi resonances.

According to the density matrix equations for atoms interacting with a phase-modulated MW field, and the small-signal approximation[29], the $\alpha$ Rabi resonance vanishes when field-atom detuning $\Delta$ is zero, and $\beta$ Rabi resonance is given as:

$$P_\beta = \frac{1}{4} \frac{m^2 \omega_m \Omega^2 \gamma_2}{[\gamma_2^2 + \Delta^2 + (\gamma_2/\gamma_1)\Omega^2]\sqrt{(\Omega^2 - 4\omega_m^2)^2 + 4\gamma_1^2 \omega_m^2}}, \qquad (2)$$

where $\gamma_1$ and $\gamma_2$ are the longitudinal and transverse relaxation rates, respectively, $\Omega$ is the Rabi frequency of the MW field, and $m$ is the modulation index.

Since the phase-modulated MW field drives the atomic population of atoms, with an amplitude of Rabi resonance signal, the optical density variation of atom vapor $\Delta OD$ is proportional to the $\beta$ Rabi resonance signal $\Delta OD \propto P_\beta$ when detuning $\Delta$ is zero.

From Equation (1), $P_\beta$ exhibits maximal amplitude when the phase modulation frequency $\omega_m = \Omega/2$. Thus, $\Omega$ is obtained from the plotted lineshape. Then, the strength of the measured

MW magnetic field strength can be calculated as $B = \frac{\hbar\Omega}{|g_J \mu_B \langle F', m_F' |J| F, m_F \rangle|}$, where $\hbar$ is the reduced Planck's constant, $g_J$ is the electron Landé g-factor, $\mu_B$ is the Bohr magneton, and $\langle F', m_F' |J| F, m_F \rangle$ is the matrix element.

The experimental setup of the microwave magnetic field detection system based on Rabi resonance is shown in Fig. 1. A distributed feedback (DFB) laser was locked onto the F = 4 - F' = 4 transition Cs D1 line (895 nm) using a saturated absorption technique. The diameter of the laser beam was 8 mm, and the optical power was 50 µW. The laser beam passed through a 1 $cm^3$ cubic glass vapor cell, filled with Cs atoms and 40 torr $N_2$. The cell was placed in 3D Helmholtz coils to compensate for the geomagnetic field and to generate a static magnetic field along the direction of the laser beam. A static magnetic field of 100 mG was applied in order to split Zeeman transitions and to isolate the magnetic field-insensitive |F=3, $m_F$=0> - |F=4, $m_F$=0> transition, which is used to generate Rabi resonance. A standard antenna was placed under the vapor cell to radiate a phase-modulated MW magnetic field. Moreover, a heating plate was placed on the cell to maintain a high enough atomic density. The heating temperature was around 55°C. The laser beam was split into two directions through a half wave plate and a polarized beam splitter. One direction was detected with a photodiode (PD) connected with a Fast Fourier Transform (FFT) spectrum analyzer; the PD detected the Rabi resonance signal of the whole laser beam in this case. Another beam was sent to an ultra-low noise complementary metal oxide semiconductor (CMOS) camera controlled by a computer, where imaging of the MW magnetic field in the range of the laser beam could be conducted. The entire experimental setup was surrounded with MW-absorbing foam to ensure an electromagnetic interference-free environment.

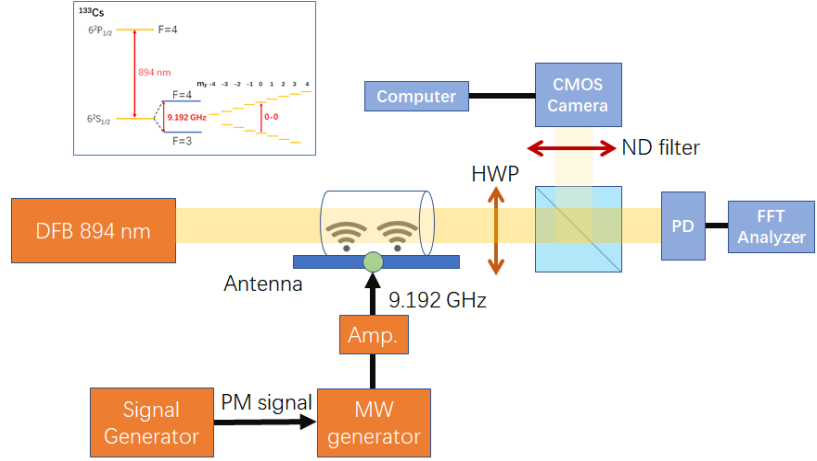

FIG. 1. Schematic of the experimental setup. The inset figure represents the energy levels of the Cs D1 line and the 0-0 transition that we used to generate the Rabi resonance signal. DFB: distributed feedback diode laser; PM: phase modulation; PD: photodiode.

The antenna used in the experiment was a simple MW coplanar waveguide which could generate a gradient MW magnetic field in free space. The MW signal was generated by a commercial low noise MW synthesizer (Agilent E8257D), the frequency being set to 9.192631770 GHz, resonant with the ground state hyperfine splitting of Cs atoms. An external signal generator was connected to the synthesizer to phase modulate the MW signal. A MW amplifier amplified the phase-modulated field with a typical gain of 35 dB. The MW power output was stabilized by a feedback loop for all measurements. MW power was detected by a Schottky diode detector and then the output compared with an ultra-stable voltage reference. Imaging of the MW magnetic field was performed with a high quantum efficiency commercial CMOS camera (PCO.Edge 4.2 LT). A lens and a density filter are placed between the cell and the camera to reduce the scale of image and avoid saturation of the CMOS.

Typical Rabi resonance signal lineshapes measured by the PD and the FFT spectrum analyzer are shown in Fig. 2. By scanning the phase modulation frequency, $\omega_m$, the Rabi resonance amplitudes could be measured. According to Equation (2), when the phase modulation frequency $\omega_m$ equals half value of the Rabi frequency of the measured MW magnetic field, the Rabi resonance signal achieves the maximum. Consequently, the Rabi frequency/strength of the measured MW magnetic field could be obtained by plotting the Rabi resonance signal lineshape.

Rabi resonance signal lineshapes with different MW input powers of the antenna are shown in Fig. 2. Each lineshape is fitted with the theoretical model. The measured Rabi frequencies clearly increase linearly with MW input power, which validates our MW magnetic field detection technique based on Rabi resonance. However, the distribution of the MW magnetic field strength in free space cannot be fully described using the Rabi resonance signal lineshape in Fig. 2. The atomic information carried in the light signal, detected with the PD, is a weighted average result of the total MW field covered by the laser beam.

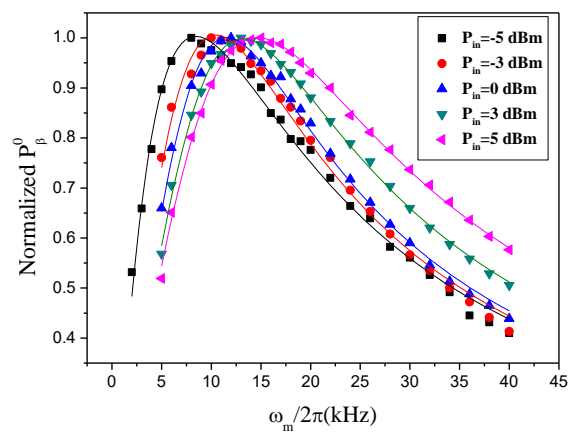

FIG. 2. Measured Rabi resonance signal lineshapes of various MW input powers.

To obtain the MW magnetic field distribution, a high quantum efficiency CMOS camera was used to capture the absorption imaging of the probe laser. The absorption of the laser was indicated by the variation in optical density of the atomic vapor, $\Delta OD = -\ln(I_{mw}/I_0)$, where $I_{mw}$ is the transmitted optical intensity on application of the MW signal, and $I_0$ is the incident optical intensity without the MW field. In each measurement, several images of $I_{mw}$ and $I_0$ were captured continuously and averaged to mitigate the noise introduced by spatial intensity variation of the laser and various other experimental errors (e.g., cell temperature fluctuations and phase noise).

Figure 3(a) shows the distribution of the gradient MW magnetic field radiated from the coplanar waveguide which is presented with the variation of optical density $\Delta OD$. The input power of the MW is 5 dBm, and the diameter of the laser beam is 8 mm. Figure 3(b) shows the simulation field distribution calculated using the simulation software ANSYS HFSS. The simulation result is largely achieved using the finite element method. Simulation complexity is based on mesh size, consequently, there is a trade-off between simulation accuracy and efficiency, i.e., computing time. Because of this trade-off, the simulation result exhibits several discrete patterns. The distribution of the MW field, captured with the camera, could avoid such an issue. Compared with the simulation result, spatial resolution of the practical experimental data for the MW magnetic field distribution is remarkably better.

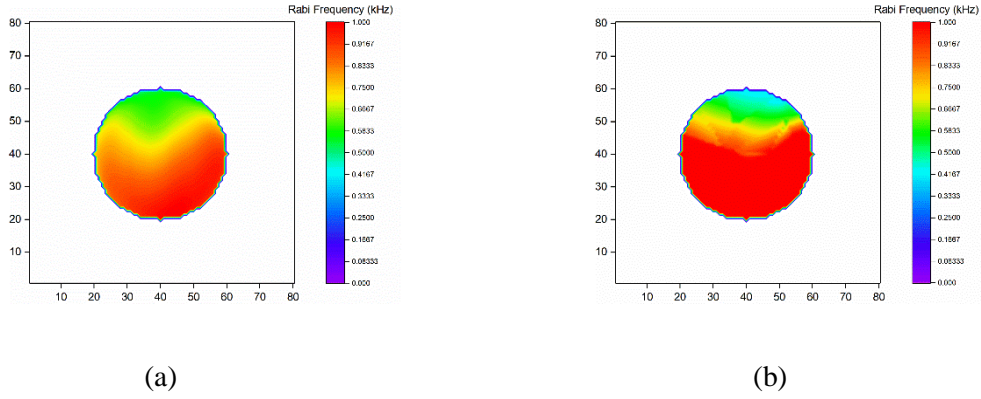

(a)    (b)

FIG. 3. Measured (a) and simulation (b) results of the MW magnetic field radiated from a standard string line. The input MW power is 5 dBm. The diameter of the laser beam is 8 mm and demagnified to 4 mm by a lens placed before the camera. The XY axis is scaled by $100\ \mu m$. The cell temperature is 55°C. The MW distributions in both images are normalized.

The MW magnetic field strength distribution can be measured quantitively when the Cs atoms are resonant with a phase modulated MW field. Variation in optical density is given as $\Delta OD_k = -\ln(I_k/I_0)$, where $I_k$ presents the transmitted optical intensity variation caused by $\beta$ Rabi resonance when a phase-modulated MW field is applied, and $f_k$ is the phase modulation frequency. Unlike the continuous signal detection with the PD and spectrum analyzer, the optical intensity variation caused by $\beta$ Rabi resonance cannot be extracted individually by the camera. To extract the optical intensity variation caused by $\beta$ Rabi resonance, considering Nyquist-Shannon sampling theorem, one way is to capture numbers of images for each PM frequency $f_k$ with an exposure time shorter than $1/2f_k$, and then we could extract the oscillation amplitude of Rabi resonance for all the pixels in the camera. However, this method requires a very short exposure time which is very difficult to realize with our camera. Note that only the relative signal amplitudes for different PM frequencies are needed to plot lineshape, consequently, we could use the variance value of numbers of images captured to present the relative Rabi resonance signal for different $f_k$. A typical capture sequence is shown in the figure 4. Numbers of images are taken for each phase modulation frequency $f_k$ with the same time interval and exposure time (typically 1 ms for time interval and 100 μs for exposure time), and then the variance of these images is calculated to present the relative transmitted optical intensity variation caused by Rabi resonance for each $f_k$. Scanning the variance of images sequence with different phase modulation frequencies, the PM frequencies correspond to the maximal $\Delta OD_k$ for all the pixels could be obtained.

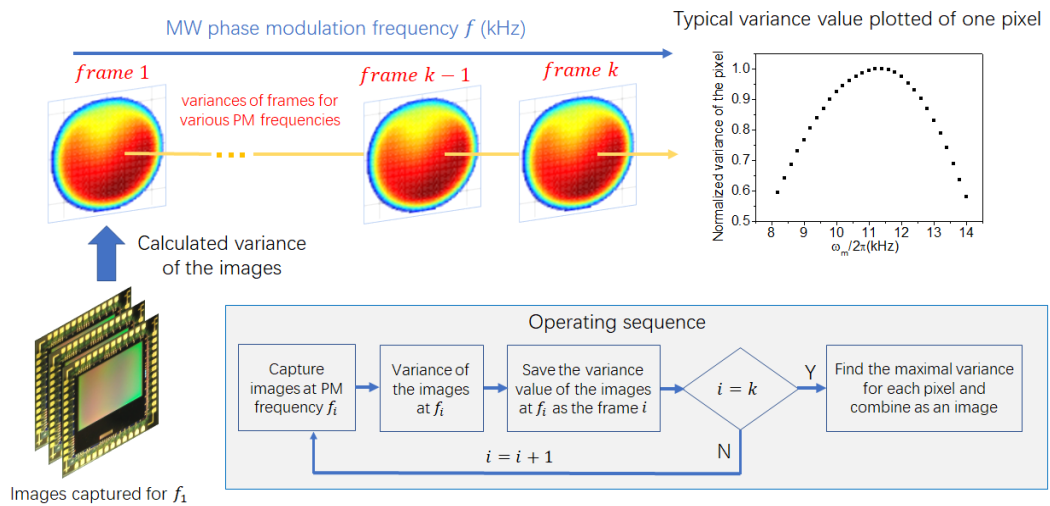

FIG. 4. Sequence of frames captured by the CMOS camera for different MW phase modulation frequencies, $f_k$. The inset figure is a typical variance value as a function of PM frequency at the same pixel in a sequence of frames. The flow chart of the operating sequence is also shown. The CMOS sensor pictures are extracted from the article "active pixel sensor" of Wikipedia.org.

Figure 5 shows the measured Rabi frequencies of the MW magnetic field distribution. Using data from the sequence of measured Rabi frequencies, Fig. 6 shows the measured Rabi frequencies of a camera pixel (marked with cross in Fig. 5) versus the input MW power. The pixel size of the camera used in the experiment was 6.5 μm × 6.5 μm. The measurement result of the camera pixel also shows a linear relationship between the measured Rabi frequencies and input MW power. The slope obtained from the camera pixel is the same as that obtained from the photodiode, confirming that our simple imagining technique has the same robust MW detection capability as conventional Rabi resonance-based sensing.

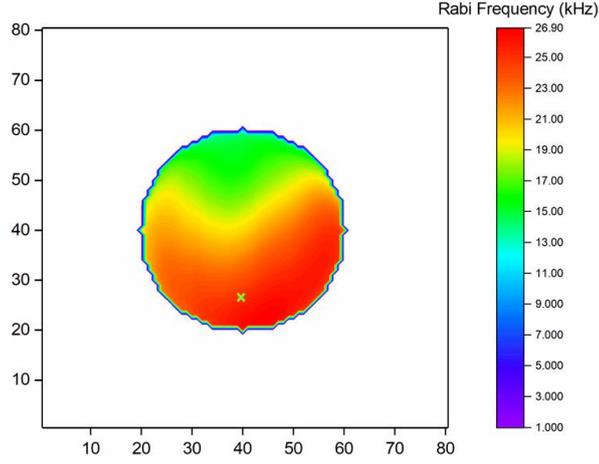

FIG. 5. Measured Rabi frequencies of the MW magnetic field in the range of the probe laser beam. The input MW power is 5 dBm. The unit of Rabi frequency is kHz.

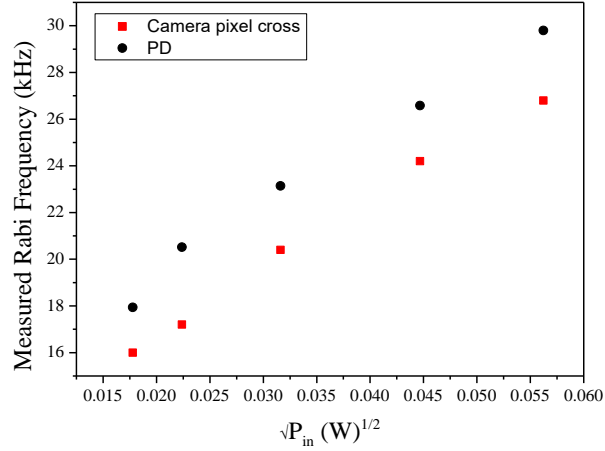

FIG. 6. Measured Rabi frequencies of a camera pixel in the image captured with camera (red) and with PD (black) as a function of various input MW powers.

In summary, we have demonstrated an imaging technique for a MW magnetic field based on Rabi resonance with a Cs-buffer gas vapor cell. By scanning the frames captured for various phase modulation frequencies with a camera, the Rabi resonance signal lineshape could be plotted for each pixel in the range of the probe laser. The strength of MW magnetic field distribution could be measured with extracting the phase modulation frequency corresponding to the maximum Rabi resonance signal value for all pixels. We were able to measure both the conventional Rabi resonance signal and the image of MW magnetic field. The experimental system consisted mainly of a frequency-locked DFB laser, a vapor cell filled with alkali atoms and buffer gas, a PD and a low noise CMOS camera. The image of the MW magnetic field captured using our technique agrees well with the simulation imaging result by the finite-element method, and even improves upon the latter's imaging quality. The probe laser beam diameter was 8 mm in our system, which determined the imaging range and resolution. The beam size could be enlarged or narrowed depending on requirements. The measurements result detected with the PD and the camera in our experimental system show non-zero field strength when the input MW power is zero (Fig. 7). This is probably a residual frequency shift due to modulation and atomic parameters and the non-linear effect in weak field sensing. A similar effect has been observed in MW electric/magnetic field measurements.[16-17, 30-31] In future work, a detailed characterization of optimal experimental parameters and the frequency shift effect of this MW magnetic field imaging technique should be investigated. Since MW magnetic field imaging have also performed with Rabi oscillation method[7,10,26], it would be very interesting to compare the imaging technique based on Rabi resonance with that based on Rabi oscillation. The simple architecture of our experimental setup holds great potential to develop a compact/miniature MW magnetic field system by combing a microfabricated vapor cell.

We thank Dr. Pengfei Wang of Wuhan Institute of Physics and Mathematics for fruitful discussions on MW magnetic field simulations. This work is funded in part by the National Key Research and Development Program of China under Grant 2018YFF0212406, 2016YFF0200104 and the Beijing Natural Science Foundation under Grant 4182078.